\documentclass[12pt]{iopart}

\usepackage{graphicx}
\usepackage{multirow}
\usepackage{cite}
\newcommand{\ket}[1]{\ensuremath{\left|{#1}\right\rangle}}
\newcommand{\bra}[1]{\ensuremath{\left\langle{#1}\right |}}
\newcommand{\oper}[1]{\mathbf{\mathsf{#1}}}
\usepackage{color}

\begin{document}

\title[Frequency-Bin Entanglement]{Frequency-Bin Entanglement of Ultra-Narrow Band Non-Degenerate Photon Pairs}

\author{Daniel Riel\"ander$^{1,*}$, Andreas Lenhard$^{1,*}$, Osvaldo Jimenez$^1$, Alejandro M\'{a}ttar$^1$, Daniel Cavalcanti$^1$, Margherita Mazzera$^1$, Antonio Ac\'{i}n$^{1,2}$ and Hugues de Riedmatten$^{1,2}$}
\address{$^1$ ICFO-Institut de Ciencies Fotoniques, The Barcelona Institute of Science and Technology, 08860 Castelldefels (Barcelona), Spain}
\address{$^2$ ICREA-Instituci\'{o} Catalana de Recerca i Estudis Avan\c{c}ats, 08015 Barcelona, Spain}
\address{$^*$ These authors contributed equally}
\ead{margherita.mazzera@icfo.eu}

\begin{abstract}
We demonstrate frequency-bin entanglement between ultra-narrowband photons generated by cavity enhanced spontaneous parametric down conversion. Our source generates photon pairs in widely non-degenerate discrete frequency modes, with one photon resonant with a quantum memory material based on praseodymium doped crystals and the other photon at telecom wavelengths. Correlations between the frequency modes are analyzed using phase modulators and narrowband filters before detection. We show high-visibility two photon interference between the frequency modes, allowing us to infer a coherent superposition of the modes. We develop a model describing the state that we create and use it to estimate optimal measurements to achieve a violation of the Clauser-Horne (CH) Bell inequality under realistic assumptions. With these settings we perform a Bell test and show a significant violation of the CH inequality, thus proving the entanglement of the photons. Finally we demonstrate the compatibility with a quantum memory material by using a spectral hole in the praseodymium (Pr) doped crystal as spectral filter for measuring high-visibility two-photon interference. This demonstrates the feasibility of combining frequency-bin entangled photon pairs with Pr-based solid state quantum memories.
\end{abstract}

\section{Introduction}
Photonic entanglement plays an important role in quantum information science, allowing the distribution of quantum correlations over long distances. It is therefore a crucial resource for several applications including quantum key distribution and quantum teleportation \cite{Gisin2007}. Photonic entanglement has been demonstrated with several processes, the most common being spontaneous parametric down conversion (SPDC). Various types of encoding have been investigated, including polarization, energy-time, time-bin and path encoding \cite{Tittel2001}. In recent years, several experiments have also investigated entanglement between frequency modes of the photons, so called frequency-bin entanglement. The sum of the frequencies of twin photons produced in an SPDC process equals the pump photon frequency. One can then define frequency bins in the spectra of the SPDC photons where one bin of the signal is correlated with the corresponding bin from the idler photon. In this context the entanglement of the photons in the frequency-bin degree of freedom has been shown \cite{Olislager2010}, and several related works have been reported \cite{Ramelow2009, Olislager2012, Bernhard2013,Xie2015,Reimer2016,Jin2016}.

For applications towards quantum networks \cite{Kimble2008} and long distance quantum communications using quantum repeaters \cite{Sangouard2011}, photons should interact strongly with material systems able to store their quantum states, i.e. photonic quantum memories \cite{Lvovsky2009,Bussieres2013,Afzelius2015}. In most cases, this requires the generation of narrowband photon pairs. Significant progress has been realized in recent years in the generation of narrow-band, memory compatible photon pairs using spontaneous parametric down conversion in non-linear crystals \cite{Bao2008,Wolfgramm2011,Foertsch2013,Fekete2013,Wang2015,Rielander2016} or Raman processes in cold atomic systems \cite{Sangouard2011,Guo2017} or rare-earth doped crystals \cite{Ferguson2016,Kutluer2017,Laplane2017}. Demonstrations of entanglement between narrow-band photon pairs have also been reported \cite{Bao2008,Clausen2011,Saglamyurek2011,Guo2017}. Time-bin or energy-time encoding are well suited for long distance fiber transmission. The analysis of such type of entanglement is usually done with the help of interferometers in a configuration introduced by Franson \cite{Franson1991}. A technical obstacle for narrowband photons with a long coherence time is that interferometers with very long path lengths are necessary. 

In this paper, we investigate the entanglement of narrow-band quantum-memory-compatible photon pairs created by SPDC in the frequency domain.  
Our work is inspired by the previous works of Olislager and coworkers \cite{Olislager2010, Olislager2012}. In these works photon pairs with broad spectra were generated and the frequency-bins were defined by tunable filters. In contrast, we developed an SPDC source based on an optical parametric oscillator (OPO) operated below its oscillation threshold \cite{Fekete2013, Rielander2016}. Our source is highly non-degenerate, with one photon at 606~nm resonant with Pr doped crystals, while the other photon is at the telecommunication wavelength of 1436~nm. The double resonant feedback in the OPO implies that the photons can only be created resonant to a few modes of the cavity. Hence in our case the frequency-bins are generated intrinsically, which, to our knowledge, was not the case in previous demonstrations. 

The entanglement between the optical modes is certified through a violation of the Clauser-Horne Bell inequality \cite{Clauser1974b}. Our measurements are implemented by electro-optic modulators to mix the frequency bins and spectral filters to select particular frequency modes. We demonstrate high visibility two-photon interference showing coherent superposition of the frequency modes. We develop a theoretical model using the photon spectrum as input that reproduces very well our measured data. Contrary to previous demonstration of frequency-bin entanglement \cite{Olislager2010, Olislager2012}, our photons exhibit a spectrum with a small number of frequency modes with different amplitude, leading to a non-maximally high-dimensional entangled state. This prevented us to use some of the assumptions done in \cite{Olislager2010}. We thus use our model to infer the best setting within our implementation to violate the inequality and show experimentally a  significant violation.

In the context of storing frequency-bin entanglement in quantum memories, rare-earth doped crystals are well suited because of their large inhomogeneous broadening of the optical transition that could be used to create frequency multiplexed quantum memories \cite{Sinclair2014,Saglamyurek2016}. Our source is compatible with praseodymium doped crystals, which have demonstrated excellent properties as quantum memories, including long storage times \cite{Heinze2013}, high-efficiency \cite{Hedges2010}, and spin-wave storage of single photons \cite{Seri2017} and of time-bin qubits at the single photon level \cite{Gundogan2015}. 

The paper is organized as follows. In section~\ref{sec:model} we describe our theoretical model. In section~\ref{sec:exp}, we describe the experimental setup and the experiments performed, including the demonstration of high-visibility two-photon interference and the violation of the Bell inequality. In section~\ref{sec:QM} we describe the two-photon interference experiments using a Pr doped crystal as spectral filter, with the goal of showing that our source is compatible with a quantum memory material. Finally, in section~\ref{sec:concl} we give a conclusion and an outlook for future experiments.

\section{Model of Frequency-Bin Entanglement}\label{sec:model}
We report on the entanglement properties of pairs of photons generated in a cavity enhanced SPDC. In this process, a pump photon of frequency $\omega_p$ enters an optical cavity where the non-linear interaction with a $\chi^{(2)}$ crystal generates pairs of signal ($\omega_s$) and idler ($\omega_i$) photons satisfying energy conservation $\omega_p=\omega_s+\omega_i$. The bi-photon state can be  described (to first order) as 
\begin{eqnarray}
	\ket{\Psi}=\int d\omega_s\int d\omega_i f_{c}(\omega_i,\omega_s)\ket{\omega_s}\ket{\omega_i}. \label{eq:biwfc}
\end{eqnarray}
 We focus on the case where both, the signal and idler photons are resonant with the cavity but are not degenerate. The joint spectral amplitude of the cavity enhanced emission is $f_{c}(\omega_i,\omega_s)$, consisting of a sequence of equally high peaks of width $\delta\omega$ separated by $\Delta_{FSR}$, the cavity Free Spectral Range (FSR). This peaked structure is modulated by the joint spectral amplitude of SPDC in the case without cavity. The explicit form of $f_c(\omega_s,\omega_i)$ is derived in \cite{Jeronimo-Moreno2010}.

As we will see in the next sections, our experimental parameters are chosen such that the widths of the peaks in the spectrum, $\delta\omega$, are two orders of magnitude smaller than the FSR, $\Delta_{FSR}$, which makes each peak perfectly distinguishable and then orthogonal to the others. This allows considering each peak as one effective mode or frequency-bin and define a discrete basis for each of the photons $\{\ket{n_{s(i)}}\}$. Consequently we can rewrite Eq.~\ref{eq:biwfc} as
\begin{eqnarray}
	\ket{\Psi}=\sum_n f_n \ket{n_s}\ket{n_i}. \label{eq:biwfd}
\end{eqnarray}

The state of Eq.~\ref{eq:biwfd} is naturally frequency-bin entangled. In order to reveal this entanglement we need to manipulate and measure the frequency modes, which we do using electro-optic modulators (EOMs) and narrow-band frequency filters. 

EOMs are optical devices, signal-controlled by a radio frequency $\Omega$ of amplitude $c$ and phase $\gamma$. When applied to a monochromatic frequency state $\ket{\omega}$ the action of the EOM can be described by the unitary transformation
\begin{eqnarray}
	\oper{U}(c,\gamma)\ket{\omega}=\sum_{n} U_n(c,\gamma)\ket{\omega+n\Omega},
\end{eqnarray}
where $U_n(c,\gamma)=J_n(c)e^{in(\gamma-\frac{\pi}{2})}$ and $J_n(c)$ are the $n$-th order Bessel functions of the first kind. Then the effect of the EOM is to create a superposition between states which are displaced by integer multiples of $\Omega$ in the frequency space. By making $\Omega$ equal to the FSR one can use the EOM to implement a unitary transformation $\oper{U}$ on the discrete basis $\left\{\ket{n_s(i)}\right\}$. 

In the experiment we are interested in measuring coincidences between signal and idler photon of a pair after they each passed an EOM and a spectral filter selecting a single frequency-bin only. The state of the pair before filtering can thus be written as:
\begin{eqnarray}
	\ket{\tilde{\Psi}}=\oper{U}_A\otimes\oper{U}_B\ket{\Psi}=\sum_n\sum_k\sum_m{U_k(c_i,\alpha)U_m(c_s,\beta)f_n\ket{n+k}\ket{-n+m}}\label{eqn:pair_eom_state}\\
	U_k(c_i,\alpha)U_m(c_s,\beta)= J_k(c_i)J_m(c_s)\exp{\left[i\left(k\alpha+m\beta-\pi/2\left(k+m\right)\right)\right]}\nonumber
\end{eqnarray}
If the spectral filters select the frequency bins $\ket{a}_s$ and $\ket{b}_i$, we can define the mode number separation $d=a-b$. The coincidence probability can then be calculated by:
\begin{eqnarray}
	P_d\left(c_i,\alpha,c_s,\beta;l\right) = \left| \left\langle l \right| \left\langle -l+d \right| \left.\tilde{\Psi}\right\rangle\right|^2\\
	\left\langle l \right| \left\langle -l+d \right| \left.\tilde{\Psi} \right\rangle = \sum_n{ \sum_k{ \sum_m{ U_k\left(c_i,\alpha\right) U_m\left(c_s,\beta\right) f_n \left\langle l\right. \left|n+k\right\rangle \left\langle -l+d\right. \left|-n+m\right\rangle}}}\nonumber
\end{eqnarray}
which, due to orthogonality and normalization of the modes, results in
\begin{eqnarray}
		P_d\left(c_i,\alpha,c_s,\beta;l\right) &=& \left|\sum_n{ \sum_k{  J_k\left(c_i\right) J_{d-k}\left(c_s\right) \exp{\left[ i\left(k\alpha + \beta\left(d-k\right) -\pi/2d \right) \right]} }}\right. \nonumber\\
	&\cdot&\left. f_n \left\langle l\right. \left|n+k\right\rangle \left\langle -l+d\right. \left|-n-k+d\right\rangle\right|^2 \label{eqn:coinc_probability}
\end{eqnarray}

In general, EOMs in conjunction with spectral filters can be used to implement a variety of measurement settings, in this case projectors
\begin{eqnarray}	
	\Pi_n(c,\gamma)=\oper{U}(c,\gamma)\ket{n}\bra{n}\oper{U}^{\dagger}(c,\gamma),
\end{eqnarray}
from where one obtains the outcome probabilities
\begin{eqnarray}
	P(ab|xy)={\rm Tr}[\Pi_a^s(x)\otimes \Pi_b^i(y)  \rho],
\end{eqnarray}
where we abbreviate the notation by using $x(y)$ for the amplitude and phase of EOM, corresponding to the measurement choices, and $a(b)$ for the measurement outcomes on signal (idler) mode.

In an experiment with photons, the probability outcome is well approximated by photon coincidence counts $C(ab|xy)$ normalized by the total number of events $\sum_{ab}C(ab|xy)$, that is
\begin{eqnarray}
	P(ab|xy)=\frac{C(ab|xy)}{\sum_{ab}C(ab|xy)} \label{prob}
\end{eqnarray}

\begin{figure}[htb]
	\centering
		\includegraphics[width=0.95\textwidth]{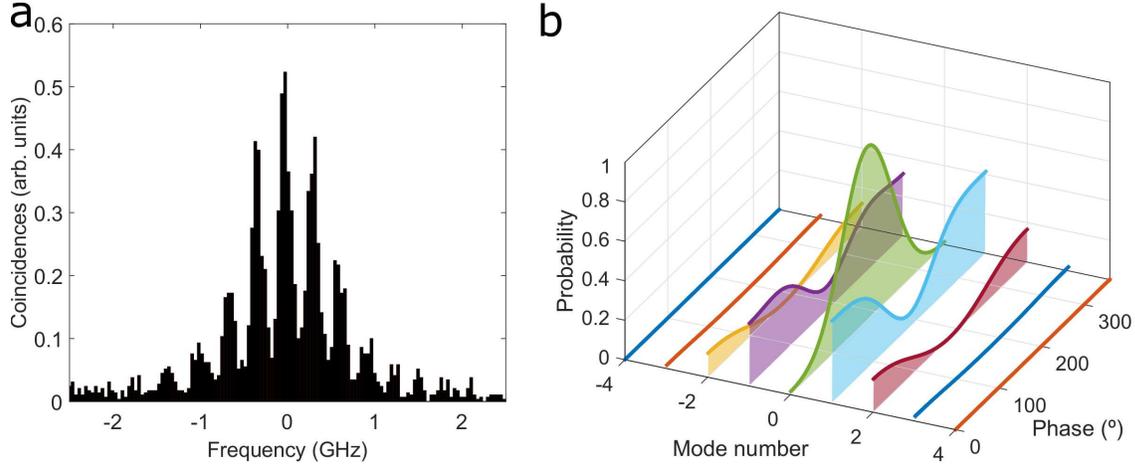}
	\caption{a) Spectral distribution of coincidences, measured without signal filter by scanning the idler filter cavity. b) Simulation of frequency-bin interference. The graph shows the spectral distribution of coincidences for a variation of phase difference between the modulators. Modulation indices $c_s=1.3$, $c_i=1.36$.}
	\label{fig:SpecModel}
\end{figure}

In practice we don't have access to all modes $\ket{a,b}$. In our experiments we can select only the central mode in the signal arm and only three modes in the idler, i.e. the central and the next neighboring modes, as we have no reference laser available for actively stabilizing the filter frequency elsewhere. However, as described by the Bessel functions, for small modulation amplitude, the effect of the EOM for $|d|>1$ is negligible. Hence we make the assumption: $\sum_{ab}C(ab|xy)\approx\sum_{|a-b|\leq1}{C(ab|xy)}$ which results in $\sum_{|a-b|\leq1}{P(a,b|xy)}\approx1$. For stronger modulation amplitudes $c_i,c_s$ this assumption still holds for a range of phases around $|\alpha-\beta|\approx\pi$, as seen in Fig.~\ref{fig:SpecModel}b. In the following section we use our model to provide an estimation about the error that this assumption for the experiment introduces.

Finally, by combining probabilities as we described above we can compute the Clauser-Horne (CH) inequality \cite{Clauser1974b}
\begin{eqnarray}\label{CH}
P(00|x_0 y_0)+P(00|x_0 y_1)+P(00|x_1y_0)-P(00|x_1y_1)  \nonumber \\ -P(0|x_0)-P(0|y_0)\leq 0,
\end{eqnarray}

where $x_i$ and $y_j$ refer to the different measurement choices. We identify the marginal probabilities with $P(0|B) = \sum_a{P(a,0|xy)}$ and $P(0|A) = \sum_b{P(0,b|xy)}$. However, since we can only detect in the central mode in the idler arm, we make the assumption that $P(0|x_0)=kP(0|y_0)$, where $k$ is a constant. Assuming symmetric spectra for the signal and idler photons (which is justified by energy conservation) before the EOMs and equal modulation indices in the signal and idler EOM, we expect $k=1$. However, in practice small variations of the modulation indices may arise. To take this effect into account, we use the model described above to estimate $k$ with our experimental parameters (see section \ref{sec:bell}). 

Since the normalization factor is common to all the involved probabilities, we can rewrite the inequality in terms only of number of coincidences.
\begin{eqnarray}
C(00|x_0 y_0)+C(00|x_0 y_1)+C(00|x_1y_0)-C(00|x_1y_1) \nonumber \\ - C(0|x_0) - C(0|y_0) \leq 0 
\end{eqnarray}
or 
\begin{eqnarray}\label{eqn:bell_inequality}
S = 2\frac{C(00|x_0 y_0)+C(00|x_0 y_1)+C(00|x_1y_0)-C(00|x_1y_1)}{C(0|A) (1+k)}\leq 2
\end{eqnarray}

Using a Matlab code we searched for the settings to achieve a high interference contrast or a significant violation of the Bell inequality (\ref{CH}). 
Figure~\ref{fig:SpecModel}b shows an example of such a simulation. As input we use the spectral distribution of the modes (Fig.~\ref{fig:SpecModel}a) that is specific for our system. The other input parameters are the modulation indices for signal and idler and a phase offset for idler ($\alpha_i=0$). We keep one mode ($\ket{0_i}$) fixed as reference for heralding. The result is the phase dependent distribution of coincidences for each mode under consideration. The normalization is realized by building the sum over mode separations $\left|d\right|\leq4$.

\section{Experiment}\label{sec:exp}
We generate photon pairs via spontaneous parametric down conversion (SPDC) in an optical parametric oscillator (OPO). The details of the device can be found elsewhere \cite{Fekete2013, Rielander2016}. We use a narrowband pump laser at 426~nm to generate idler photons at 1436~nm (telecommunications E-band) and signal photons at 606~nm. The source design allows us to generate signal photons compatible with solid state quantum memories based on praseodymium doped crystals \cite{Rielander2014,Seri2017}. The spectrum of the photons consists of several peaks, formed by longitudinal modes of the OPO cavity (see Fig.~\ref{fig:SpecModel}a and \cite{Rielander2016}). Hence, these modes are separated by the cavity FSR, which we measured as $\Delta_{FSR}=423.66(6)$~MHz. The measured biphoton bandwidth is 2.8~MHz. In total we can identify up to eight modes contributing to the spectrum with different intensities. Our source setup further contains tunable Fabry-P\'{e}rot cavities (FC) which we use to select a single mode of the spectrum. These filters are actively stabilized to a reference laser. Fulfilling this preconditions our source is well suited to demonstrate frequency-bin entanglement. 
\begin{figure*}[htb]
	\centering
		\includegraphics[width=0.95\textwidth]{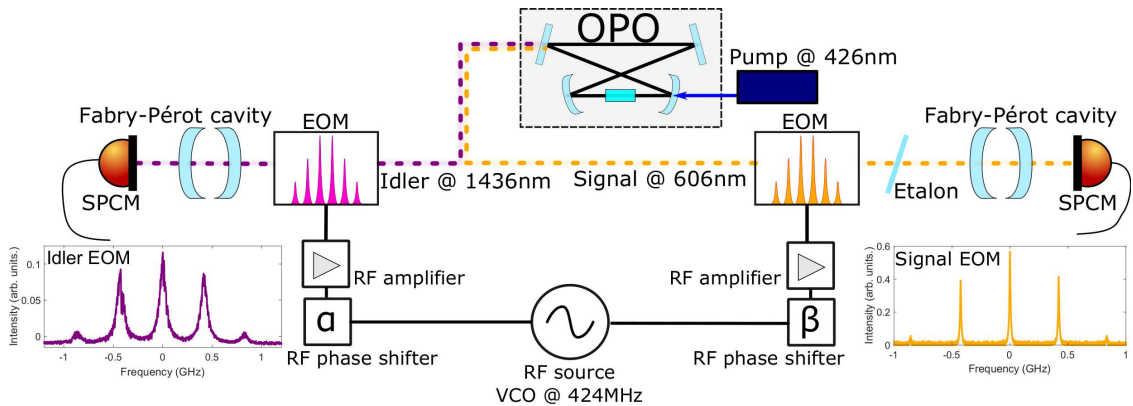}
	\caption{Experimental setup. Details are explained in the text. The insets show the creation of sidebands by the EOMs on the single-mode reference lasers.}
	\label{fig:Setup}
\end{figure*}

To analyze the frequency-bin entanglement we include electro-optic modulators (EOMs) in our setup as illustrated in Fig.~\ref{fig:Setup}. Each EOM is placed behind the source, before any spectral filter. We use a voltage controlled oscillator (VCO), tuned to 423.7~MHz ($\approx\Delta_{FSR}$), as master oscillator. We split its radio frequency (RF) signal in two parts and feed it to the following components which are identical for the signal and idler branch. The further path for the RF signal contains a phase shifter, a variable attenuator and a high power amplifier to drive the EOM. Hence we can set the phase and power for each EOM individually, preserving coherence between the two EOMs. The effect of the EOMs on the optical spectrum is the creation of sidebands, as can be seen in the insets of Fig.~\ref{fig:Setup}, where we send a single mode, continuous wave laser beam through the EOMs and measure its spectrum by scanning the length of the filter cavities. It is worth mentioning that the widths of the peaks in this figure are limited by the resolution of the FCs, which is different for signal and idler. From the ratio of the peak height, we infer a modulation index $c_i=1.36$ for idler and $c_s=1.30$ for signal. The modulation index can be adjusted by the RF power sent to the EOMs. We found a linear relation between the modulation index and the RF amplitude. We use measurements of the RF amplitude at each data point to keep the modulation index constant.

The stabilization procedure of the FC limits our access to the modes during the measurements. The cavity for idler can be stabilized to the modes $\ket{n_i}= \ket{-1_i},\ket{0_i},\ket{+1_i}$, while in the signal arm we can only select the central mode $\ket{n_s} = \ket{0_s}$. In contrast to our model or simulation, for the normalization of the coincidence rates we can only consider these modes. Hence we assume that higher mode separation can be neglected if the modulation is chosen low enough. This effect is illustrated in Fig.~\ref{fig:Fringes}a. The black solid line is the sum of all modes resulting from the simulation, while for the dashed black line the summation was restricted to the three central modes (as in the experiment). The gray dots in that figure are the sums of the experimental data points and overlap well with the dashed line. At phases around the maximum for $d=0$ the coincidence rate for $\left|d\right|>1$ is negligible, which can be seen in the good overlap between the black lines. We chose four data points in this range (marked with circles in Fig.~\ref{fig:Fringes}a) to scale the simulation results from probabilities to coincidences. For direct comparison all solid lines are normalized taking all modes into account, while the dashed lines follow the experimental normalization scheme with three modes only. For the results with $d=0$ we see that the difference between the two methods is smaller than the experimental error bars.

\subsection{Frequency-Bin Two-Photon Interference}\label{sec:interfer}
\begin{figure}[htb]
	\centering
		\includegraphics[width=0.95\textwidth]{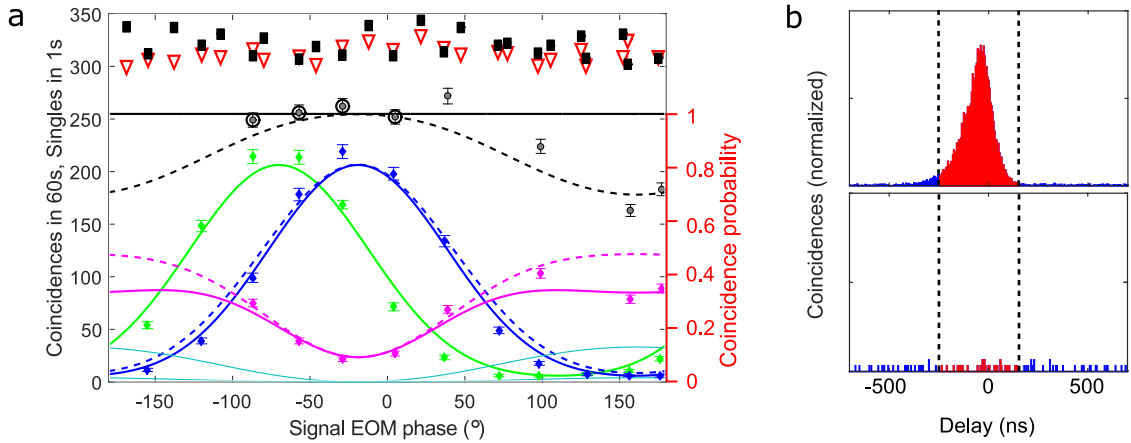}
	\caption{Results of the correlation measurements. a) The empty red triangles and the filled black squares show the measured single rates for signal and idler, respectively. The green points show the coincidence rate for idler phase set to $\alpha=0^\circ$ while for the blue points $\alpha=51^\circ$. Both signal and idler filter cavities were set to mode $\left|0_{s,i}\right\rangle$. The solid lines show fits to the data. The magenta points show coincidences with $\alpha=51^\circ$ and the idler filter selecting mode $\left|+1_i\right\rangle$. The solid lines show the results from our simulations. The cyan colored lines are simulations for higher order modes, not accessible in the experiment.The dashed blue line corresponds to the solid blue simulation, normalized by the three central modes only, as in the experiment. b) Shows the correlation histograms for minimum (bottom, $\beta=178^\circ$, $\alpha=51^\circ$) and maximum (top, $\beta=-29^\circ$, $\alpha=51^\circ$). To determine the coincidence rate we integrate over an 400~ns interval, indicated by the vertical dashed lines.}
	\label{fig:Fringes}
\end{figure}

As the RF is matched to the cavity FSR, the side bands of one mode will overlap with the neighboring modes. This creates the superposition of frequency bins necessary to show interference.
The phase dependence of the coincidence rate is described by Bessel functions and depends on the modulation indices (see Eq.~\ref{eqn:coinc_probability}). With the help of our model we decided to use equal depths in the range $1\leq c_{s},c_{i}\leq 1.4$ as we expect a rather high interference contrast for the coincidence rate with these settings. Examples are shown in Fig.~\ref{fig:Fringes}b. We first set the phase of the idler EOM to $\alpha_0=0^\circ$ and varied the signal EOM phase only (green curve in Fig.~\ref{fig:Fringes}). Next we set the idler EOM phase to $\alpha_1=51^\circ$ and repeated the measurement (blue curve in Fig.~\ref{fig:Fringes}). In both cases the filter cavities were actively stabilized to the central mode (mode $\left|0_{s,i}\right\rangle$) of the spectra, corresponding to $d=0$. For comparison, the magenta data points in the figure show the results when the idler cavity is stabilized to the side mode (mode $\left|+1_i\right\rangle$) while the signal filter is still centered (mode $\left|0_s\right\rangle$), corresponding to $d=1$.

While we observe a dependence of the coincidence rate on the phase settings, no changes in the single rates are observed. Thus, this variation of the coincidence rate induced by the change of the relative RF phase shows two-photon interference which is a signature of the coherent superposition of the frequency-bins. From the raw data, shown in Fig.~\ref{fig:Fringes}, we find an interference contrast of $V=95(4)\%$. The solid lines shown in the figure are predictions of our simulation. 

The very good overlap of the experimental data with the model is a first hint that we generate a frequency-entangled state.

\subsection{Bell Inequality}\label{sec:bell}
\begin{figure}[htb]
	\centering
		\includegraphics[width=0.90\textwidth]{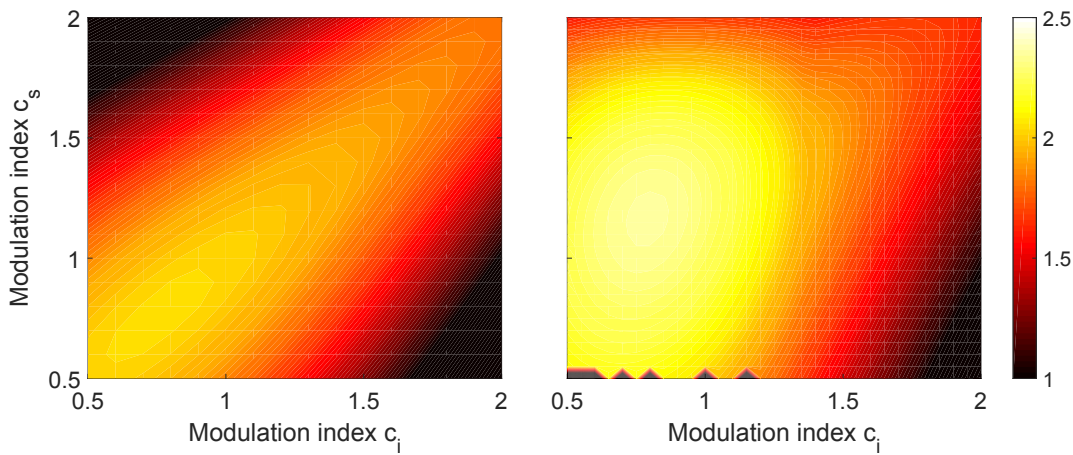}
	\caption{Simulation of highest possible S-values for different settings. In the left graph the modulation index is held constant for all four Bell settings (although it can take different values for signal and idler photons). In the right graph we assume switching the modulation indices by 0.5 and phase shift by 180$^\circ$ between the four settings. In this combination higher limits for the S-value can be achieved.}
	\label{fig:BestS}
\end{figure}

\begin{figure}[htb]
	\centering
		\includegraphics[width=0.95\textwidth]{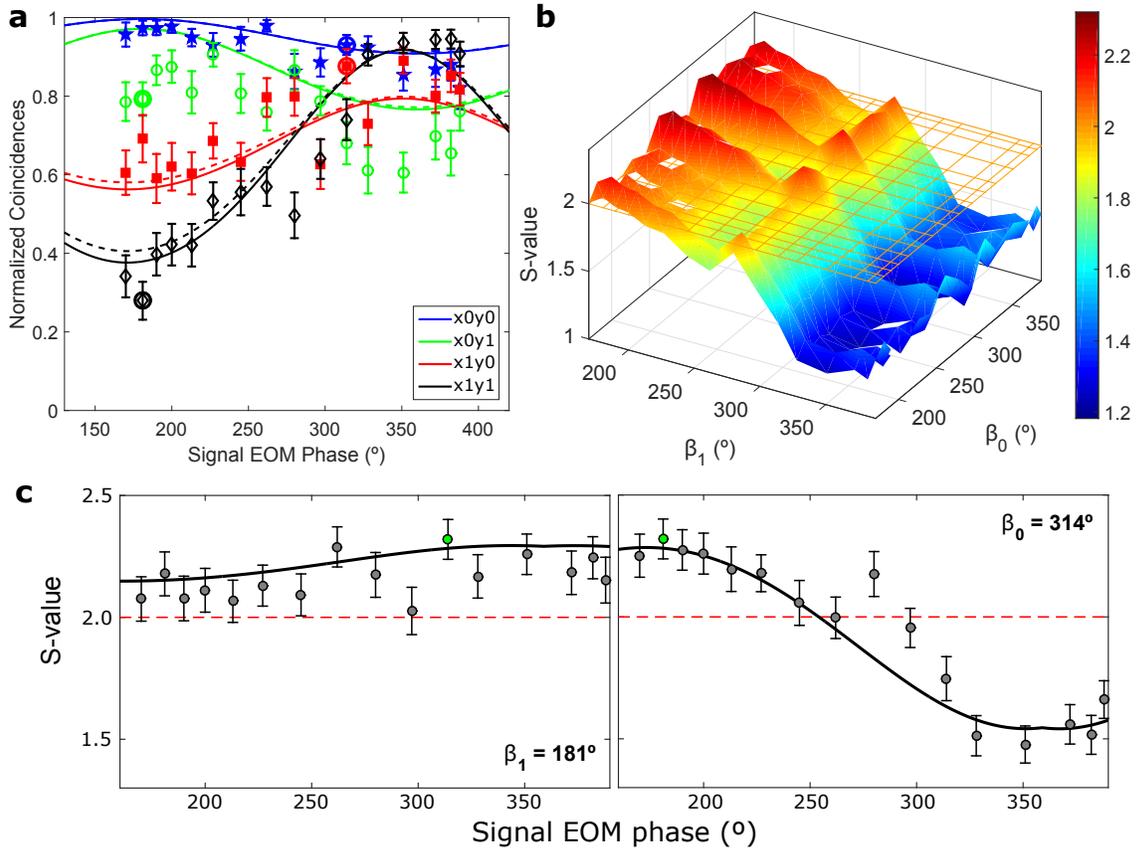}
	\caption{a)  Interference fringes for Bell Test. We performed 4310 measurement sequences for the center mode and 904 and 1230 for the side modes, each sequence integrating coincidences for 10~s. The solid lines show the simulation results. The dashed lines show the simulation, normalized by the two side modes only, according to the experimental procedure. The points marked with circles allow the highest violation of the Bell inequality. b) S-values for all allowed combinations of data points. c) Cuts through the highest point in (b) along the two phase settings. The solid lines show the simulation results, the green filled points indicate the maximum violation.}
	\label{fig:BellFringe}
\end{figure}

The results in Fig.~\ref{fig:Fringes} show a very high two photon interference contrast, offering evidence of the entanglement of the state we generate. To demonstrate entanglement without assumptions on the dimension of the created state, we then show violations of the Bell inequality (Eq.~\ref{eqn:bell_inequality}).  However, although the results in Fig.~\ref{fig:Fringes} show a very high interference contrast, we cannot reach an S-value above the classical boundary with these settings. The reason for this surprising result is that it is not possible in these conditions to find phase settings for which the side modes are zero and all the coincidences are in the central mode. This is a consequence of the fact that the created state is non-maximally entangled. We confirm this intuition by our simulation which shows that if the modulation index remains constant for the four settings (although it may be different for signal and idler photons), only a negligible violation can be achieved (see Fig.~\ref{fig:BestS} left). 
However, much more significant violations can be achieved if the modulation depth is alternated between the four settings (see Fig.~\ref{fig:BestS} right). From a technical point of view the differences in modulation during the alternation should not be too high, as we observed a reshaping of the optical beam by the EOMs depending on the thermal load (i.e. on the power of the RF), changing detection efficiency of the photons.  In the experiment a switching of the modulation index by less than 0.5 has no observable influence on the transmission of the photons to the detectors. With this limitations, the best results are expected for an alternation between a high and a low modulation index and phase shifts around $180^\circ$. The final values we choose can be found in table~\ref{tab:BellSettings}. We accordingly shift the phase of the idler RF by $182^\circ$.
\begin{table}[htb]
	\centering
		\begin{tabular}{| l | l | c | c | c | c |c |}
			\hline
			Experiment & Setting & $x_0$ & $x_1$ & $y_0$ & $y_1$ & k\\
			\hline
			\multirow{2}{*}{Fringe} & mod. idx. & 0.29 & 0.85 & 0.34 & 0.81&1.01\\
			& phase & 0 & 182 & 314 & 181&\\
			\hline
			\multirow{2}{*}{Bell Points} & mod. idx. &0.44 & 0.56 & 0.34 & 0.81 & 0.97\\
			& phase & 0 & 182 & 361 & 171&\\
			\hline
		\end{tabular}
	\caption{Bell Settings. The $k$ parameter, which accounts for small variations of the signal and idler modulation indices (see section \ref{sec:model}), is also reported.}
	\label{tab:BellSettings}
\end{table}
To calibrate the global phase offset, in a first step we measured four fringes with the four settings and varied the phase of the signal EOM over the full accessible range. We used a random number generator to decide for each trial which setting ($c_{i,j},\alpha_j,c_{s,j}$ for $j=0,1$) and phase ($\beta_k$) to use. In each trial we collected data for 10~s. The results were then added to the corresponding data point. The whole procedure was afterwards repeated with the idler filter set to the side modes ($\ket{\pm1_i}$), as necessary for the normalization. The normalized coincidences are shown in Fig.~\ref{fig:BellFringe}. 

To compare our model with the data, we measured the modulation indices for the four settings with the procedure described above and put this information into the model. The only free parameter is a global phase shift between the model and the experimental data which we fit to the best overlap. The result is plotted in Fig.~\ref{fig:BellFringe} for comparison. We additionally plotted the result in the case we normalize the simulation neglecting modes $\left|d\right|>1$ (dashed lines). The effect is smaller than the average error bar in the experiment proving the validity of our assumption. The effect of not-optimal normalization is smaller here compared to Sec.~\ref{sec:interfer} due to the lower modulation indices. Less modulation depth results in less mixing of frequency bins with neighbouring modes. Quantitatively, the two methods introduce a deviation of 7~\% for the normalization of the $x_1y_1$ setting, while the deviation is up to two orders of magnitude smaller for the other settings. As a matter of fact, the $x_1y_1$ setting has higher modulation indices, thus the contribution of the modes with $\left|d\right|>1$ is bigger. As explained in Section \ref{sec:model}, we assume that  $C(0|x_0)=kC(0|y_0)$. For the values of modulations used here (see table \ref{tab:BellSettings}), we simulate the spectrum of the signal and idler field after their respective EOM and can then estimate the value of $k$. For the measurements shown on Fig.~\ref{fig:BellFringe}, we find $k=1.01$. 

The Bell inequality is based on the comparison of four coincidence probabilities, one for each combination of settings. Figure~\ref{fig:BellFringe}a shows the result for all allowed combinations of the experimental data. As we can see in  Figures~\ref{fig:BellFringe}b and c, the violation is quite robust as there is a wide range of the measurement parameters resulting in $S>2$. However, the highest violation of $S=2.31(8)$ (calculated using $k=1.01$) is achieved with the combinations summarized in table~\ref{tab:BellSettings}. For the error estimation we here take only into account the Poissonian statistics of the photon counts, no systematic errors are considered. The errors are then combined via error propagation calculations. This results in a violation of the classical bound by four standard deviations. Our model predicts a violation of $S=2.30$ which is in perfect agreement with the experimental result. The deviation of the S-value due to the non-optimal normalization is below 1~\% and thus smaller than the statistical error of 3.5~\%. 

In addition, we repetitively measured the coincidence rate for the four settings (see table~\ref{tab:BellSettings}) to show a violation of the Bell inequality. We did these measurements with a reduced RF power in the idler EOM. Such a setting increases the stability of our FC lock system which in turn allows longer measurement times. On the other hand, we expect a lower violation of the Bell inequality ($S=2.24$). In that case, we estimate the proportionality constant between the marginals to be $k=0.97$. With the focus on these points only, a large number of coincidences could be recorded reducing the statistical error. This finally results in $S=2.21(2)$ (calculated with $k=0.97$), corresponding to a violation of 9.7 standard deviations.

\section{Compatibility with a quantum memory material}\label{sec:QM}
In order to show the compatibility of our source with a quantum memory material, we then perform experiments using a narrow spectral hole in the absorption profile of a praseodymium doped Y$_2$SiO$_5$ crystal as frequency filter to select one frequency mode in the signal arm. 
\begin{figure}[htb]
	\centering
		\includegraphics[width=0.95\textwidth]{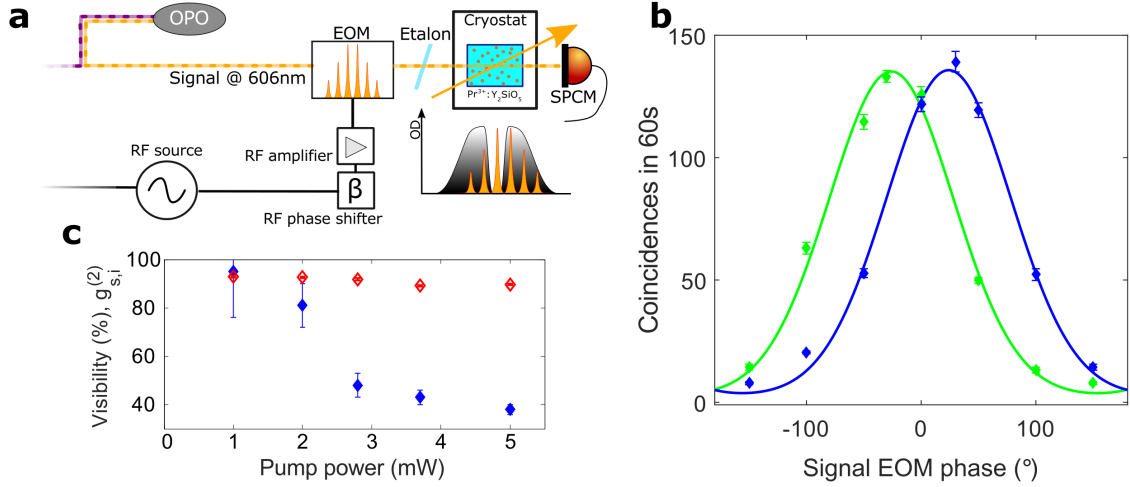}
	\caption{a) Modification of the experimental setup: The filter cavity for the signal photons was replaced by our memory setup. We create a spectral filter by spectral hole burning techniques. The resulting coincidence measurement (b) shows interference fringes. c) Measured interference raw visibility (red empty diamonds) and cross correlation value g$^{(2)}_{s,i}$ (blue filled diamonds) for different pump powers.}
	\label{fig:PitFringes}
\end{figure}
The inhomogeneous broadening of the transition at 606~nm results in total absorption of all signal modes created by our OPO. However, we can create a transmission window via spectral hole burning. With the help of a narrow band tunable laser we empty two of the three hyperfine ground states. With this technique we tailor a window of 18~MHz featuring high transmission for the mode $\left|0\right\rangle$ only. We measure coincidences between the single mode signal photons filtered in that way and the idler photons filtered to a single mode with the help of the cavity, as explained before. The coincidence rate is reduced due to increased losses in the signal arm and an additional duty cycle, necessary to periodically refresh the spectral filter hole. As shown in Fig.~\ref{fig:PitFringes}b we still observe interference fringes with high visibility.

We repeated this measurements for different pump powers of our source. Increasing the pump power increases the probability to create multiple pairs in the SPDC process. This reduces the fidelity of the heralded single photon. This effect can be observed in a decrease of the signal-idler cross correlation value g$^{(2)}_{s,i}$ (see also \cite{Rielander2016}). The result is summarized in Fig.~\ref{fig:PitFringes}c. For all available pump powers we operate in a non-classical regime. Accordingly we observe a decrease of frequency-bin interference visibility with pump power. Nevertheless, the lowest visibility for all accessible pump powers was still quite high with $V\geq89.2(3)\%$. These results demonstrate the compatibility of our source with a QM material.

However, further steps will be needed before storing frequency-bin entanglement. We demonstrated previously that we can store heralded single photons created by this source in a solid state quantum memory \cite{Rielander2014, Seri2017}, using the atomic frequency comb (AFC) technique for storage. This involves creating a periodic absorption feature within the spectral hole we used for filtering. To store a frequency-bin entangled qudit the creation of multiples AFCs on the inhomogeneous profile will be needed. This would allow us to store the photon in a superposition of frequency bins which is essential to preserve the qudit state. The EOM for analysis of the entanglement will be placed behind the memory. The creation of multiple AFCs has already been demonstrated in other materials \cite{Sinclair2014,Saglamyurek2016}. 

\section{Conclusion}\label{sec:concl}
Frequency-bins are naturally defined for photons generated by an OPO-SPDC source. We here demonstrated for the first time the entanglement of such OPO based photon pairs in the frequency-bin basis. The results can be well simulated by our generalized model based on the spectral distribution of the photons. With this model we analyzed the violation of the CH inequality that is accessible by our experiment. With our generated state, we observed that the settings used to obtain a high visibility in the two-photon interference were not suited for a strong violation of the Bell inequality, due to the shape of the photons' spectrum. Nevertheless, with optimized settings we could significantly violate the inequality and thus show the entanglement of the photons.  

In a next step we replaced the signal filter cavity with a spectral hole in a quantum memory crystal and still observe interference effects, supporting prospects for storing frequency entangled photons.

With our technique the photons are prepared in a high dimensional state. Future work will also focus in a better understanding of the dimensionality and the applications offered by this capability.

\ack
We thank Alessandro Seri for help in operating the quantum memory.
We acknowledge financial support by the ERC through the starting grant QuLIMA and the consolidator grant QITBOX, by the AXA Chair in Quantum Information Science, by the Spanish Ministry of Economy and Competitiveness (MINECO) and Fondo Europeo de Desarrollo Regional (FEDER) through Grant No. FIS2015-69535-R (MINECO/FEDER), by MINECO through Grant QIBEQI FIS2016-80773-P, by MINECO Severo Ochoa through Grant No. SEV-2015-0522, by AGAUR via 2014 SGR 1554 and 2014 SGR 875, by Fundaci\'o Privada Cellex, and by the CERCA programme of the Generalitat de Catalunya. D.C. acknowledges the Ramon y Cajal Programme. O.J.F. was supported by the Beatriu de Pin\'os fellowship (Grant No. 442 2014 BP-B 0219).

\bibliographystyle{unsrt}

\end{document}